\documentclass[12pt]{article}
\usepackage{amssymb,amsmath,epsfig}
\usepackage{graphicx}
\allowdisplaybreaks
\begin{document}
\title{\bf Study of Cosmic Evolution admitting Thermodynamic Analysis}
\author{M. Sharif \thanks{msharif.math@pu.edu.pk}~, M. Zeeshan Gul
\thanks{mzeeshangul.math@gmail.com}~ and~ Nusrat Fatima
\thanks{nusratfatimaliaqat@gmail.com}\\
Department of Mathematics and Statistics, The University of Lahore,\\
1-KM Defence Road Lahore-54000, Pakistan.}

\date{}
\maketitle

\begin{abstract}
This article examines the cosmic evolution in the framework of
symmetric teleparallel theory, characterized by the function of
non-metricity scalar $(\mathcal{Q})$. We use the e-folding number
and reconstruction method with a suitable parametrization of the
scale factor to obtain the functional form of symmetric teleparallel
theory. Using this reconstructed model, we examine the behavior of
different cosmographic parameters to demonstrate the bouncing
scenarios of the cosmos by considering the contraction and expansion
phases of cosmos before and after the bouncing point, respectively.
It is found that the null energy condition is violated which shows
that the singularity issue can be resolved in this extended
theoretical framework. Moreover, we observe that the acceleration
occurs near the bouncing point and the reconstructed model aligns
with the current cosmic expansion. Finally, we check the validity of
second law of thermodynamics in the bouncing framework of our model.
\end{abstract}
\textbf{Keywords}: Bounce models; Modified theory; Cosmological
parameters.\\
\textbf{PACS}: 64.30.+t; 04.20.Dw; 04.50.kd.

\section{Introduction}

The current accelerated expansion of the cosmos has been a
captivating and significant discovery for researchers over the past
two decades. Scientists claimed that this cosmic expansion is driven
by a mysterious force, known as dark energy (DE). Cosmologists have
been motivated to investigate the elusive nature of this enigmatic
energy. To account the puzzling features of DE, Einstein introduced
the cosmological constant into his field equations, known as
$\Lambda$CDM model. However, this model faces two primary
challenges, i.e., the coincidence problem and the fine-tuning
problem. The coincidence problem arises from the notable discrepancy
between the observed and predicted values of energy density. Another
question is why we are observing the current cosmic acceleration
when the energy densities of dark matter and DE are assumed to be
equal \cite{1}. To address these problems and unveil the cosmic
mysteries, various modified gravitational theories (MGTs) such as
curvature-based, torsion-based and non-metricity-based theories have
been developed \cite{2}-\cite{5j}. One of the well-known approaches
to discuss the dark universe is $f(\mathcal{Q})$ gravity, where the
gravitational Lagrangian involves a generic function of the
non-metricity \cite{7}.

The study of $f(\mathcal{Q})$ theory has gained significant interest
in recent times due to its consistency with observational data and
its implications for cosmology. Researchers have investigated
various aspects of this theoretical framework. Lazkoz et al \cite{8}
studied the cosmic evolution using redshift function in this theory.
Jimenez et al \cite{9} examined the cosmic perturbations in this
modified framework. Mandal and his colleagues \cite{10} examined the
energy conditions (ECs) to evaluate the viability of cosmological
models in the same theoretical context. Bajardi et al \cite{11}
utilized the Hamiltonian approach to derive the cosmic wave
function. Hassan et al \cite{12} explored the stability analysis of
wormhole solutions in the symmetric teleparallel theory. Solanki et
al \cite{13} investigated the role of bulk viscosity in the cosmic
accelerated expansion. Esposito et al \cite{14} used reconstruction
techniques to study the precise isotropic and anisotropic
cosmological solutions. Arora and Sahoo \cite{15} looked into the
cosmic evolution through the EoS parameter. Albuquerque and
Frusciante \cite{16} examined the evolution of linear perturbations
in the same theory. Sokoliuk et al \cite{17} explored the evolution
of the universe using Pantheon data sets. Khyllep et al \cite{18}
analyzed the cosmic expansion employing power-law and exponential
forms of this gravity to understand the cosmic dynamics.  The
geometry of compact stars with different considerations in
$f(\mathcal{Q})$ and $f(\mathcal{Q},T)$ theory has been studied in
\cite{19a}-\cite{19g}, where $T$ is the trace of the energy momentum
tensor (EMT). The $f(\mathcal{Q})$ gravity models in a
Friedmann-Robertson-Walker (FRW) and Bianchi type-I spacetimes
provide valuable insights into the role of non-metricity in
gravitational effects and cosmological scenarios
\cite{20}-\cite{21}.

Cosmological observations indicate that the universe originated from
a singularity, known as the big bang. However, this theory faces
several issues such as the horizon and flatness problems. To address
these problems, the inflation theory was developed which provides a
basic explanation for cosmic expansion. While the inflation theory
\cite{22} successfully resolves many issues but it fails to address
the initial singularity problem. In this regard, a viable cosmic
model called bouncing cosmology has been proposed to resolve the
initial singularity. The bounce theory describes a cyclic pattern
where the collapse of one cosmic event precedes the occurrence of a
new one. During this bouncing cosmological behavior, the Hubble
parameter shows the transitions from contraction to expansion
phases. Bounce solutions are significant in cosmology as they offer
a way to solve the initial singularity problem.

Bouncing cosmology in MGTs has attracted significant interest
because of their intriguing features. Barragan et al \cite{23}
examined bouncing cosmology in Palatini $f(R)$ gravity. Saaidi et al
\cite{24} used the Bianchi type-I spacetime to analyze cosmic
evolution in $f(R)$ framework. Jawad and Rani \cite{25} studied
cosmic evolution through a generalized ghost DE model in the same
work. Shabani and Ziaie \cite{26} investigated non-singular bouncing
solutions with perfect matter configuration in $f(R,T)$ framework.
Aly \cite{26a} studied the generalized second law of thermodynamics
in the background of Ricci DE models, examining both interacting and
non-interacting scenarios. Malik and Shamir \cite{27} explored the
bouncing cosmos in the same theory and found that specific solutions
can accommodate exotic matter. Shekh \cite{27a} used the second law
of thermodynamic to analyze the dynamical behavior of anisotropic DE
models in the framework of $f(R,G)$ theory, where $G$ is the
Gauss-Bonnet invariant. Ilyas et al \cite{28} studied the cosmic
dynamics through different $f(R)$ models and observed that these
models can resolve the initial singularity. Zubair et al \cite{29}
discussed the viability of the reconstructed cosmic models in
$f(R,T)$ theory. Bhardwaj et al \cite{30} examined cosmic evolution
using cosmographic parameters in the same framework. Lohakare et al
\cite{31} used $f(R,G)$ gravity models to analyze the bouncing
cosmos. Dimakis et al \cite{32} studied viable anisotropic solutions
for Kantowski-Sachs and Bianchi type-III spacetimes in
$f(\mathcal{Q})$ theory. The analysis of observational constraints
in MGTs has also been explored in various studies
\cite{33}-\cite{36}. Sharif et al \cite{37} studied the cosmic
bounce in non-Riemannian geometry. Sharif and his collaborators
studied the Noether symmetry approach \cite{45a}-\cite{50a},
stability of the Einstein universe \cite{51a}-\cite{53a}, dynamics
of gravitational collapse \cite{54a}-\cite{56a} and static
spherically symmetric structures \cite{57a}-\cite{60a} in $f(R,
{T}^{2})$ theory.

This manuscript provides a framework to study the cosmic evolution
with thermodynamic analysis in $f(\mathcal{Q})$ theory. The paper is
organized in the following order. In section \textbf{2}, we define
the vacuum action in $f(\mathcal{Q})$ gravity and use the
reconstruction technique to find the functional form of modified
symmetric teleparallel theory. Also, we derive the field equations
of $f(\mathcal{Q})$ gravity in the presence of FRW spacetime. The
comprehensive analysis of the bouncing universe is presented in the
section \textbf{3}. Additionally, we use the reconstructed
functional form of $f(\mathcal{Q})$ to discuss the graphical
behavior of cosmic parameters. The brief analysis of the second law
of thermodynamics is provided in section \textbf{4}. Our main
findings are summarized in the section \textbf{5}.

\section{Reconstructed $f(\mathcal{Q})$ Functional Form}

The modified action of $f(\mathcal{Q})$ gravity in vacuum is given
by
\begin{equation}\label{1}
\mathcal{S}=\frac{1}{2\kappa}\int f(\mathcal{Q}){\sqrt{-g}} d^4x,
\end{equation}
where $\kappa=1$ represents the coupling constant and $g$ denotes
the determinant of the metric tensor. In the integrand, the
non-metricity is given by (detailed calculation is given in Appendix
\textbf{X})
\begin{equation}\label{2}
\mathcal{Q}=-\mathcal{Q}_{\xi\alpha\beta}\mathcal{P}^{\xi\alpha\beta}=
-\frac{1}{4}(-\mathcal{Q}^{\xi\alpha\beta}\mathcal{Q}_{\xi\alpha\beta}
+2\mathcal{Q}^{\xi\alpha\beta}\mathcal{Q}_{\beta\xi\alpha}-2\mathcal{Q}^{\xi}
\tilde{\mathcal{Q}}_{\xi}+\mathcal{Q}^{\xi}\mathcal{Q}_{\xi}),
\end{equation}
where
\begin{equation}\label{3}
\mathcal{P}^{\xi}_{\;\alpha\beta}=-\frac{1}{2}L^{\xi}_{\;\alpha\beta}
+\frac{1}{4}(\mathcal{Q}^{\xi}-\tilde{\mathcal{Q}}^{\xi})g_{\alpha\beta}-
\frac{1}{4} \delta ^{\xi}\;(_{\alpha}\mathcal{Q}_{\beta}),
\end{equation}
and
\begin{equation}\label{4}
\mathcal{Q}_{\xi}\equiv \mathcal{Q}^{~~\alpha}_{\xi~~\alpha}, \quad
\tilde{\mathcal{Q}}_{\xi}\equiv \mathcal{Q}^{\alpha}_{~~\xi\alpha}.
\end{equation}
To derive the gravitational field equations, one can perform a
variation of the action with respect to the metric tensor. The
explicit formulation of $\delta \mathcal{Q}$ is provided in Appendix
\textbf{Y}. We consider a flat FRW metric as
\begin{equation}\label{5}
ds^2=-dt^{2}+(dx^{2}+dy^{2}+dz^{2})a^{2}(t),
\end{equation}
where $a(t)$ is the scale factor. The gravitational field equations
can be obtained as
\begin{eqnarray}\label{6}
3{H}^{2}&=&\frac{1}{2}f-6H^{2}f_\mathcal{Q},
\\\label{7}
2\dot{H}+3H^{2}&=&2f_{\mathcal{Q}}\dot{H}+2f_{\mathcal{Q}\mathcal{Q}}H
+6f_{\mathcal{Q}}H^{2}-\frac{1}{2}f.
\end{eqnarray}
Here, $f(\mathcal{Q})\equiv f$,
$f_{\mathcal{Q}}=\frac{\partial{f}}{\partial{\mathcal{Q}}}$ and dot
is the time derivative. Using Eqs.\eqref{2} and \eqref{5}, the value
of non-metricity turns out to be (details are given in Appendix
$\textbf{Z}$)
\begin{equation}\label{8}
\mathcal{Q}=6H^{2}.
\end{equation}

Since the solution of the field equations \eqref{6} and \eqref{7} is
very complicated as they contain multi-variables and their
derivatives. Thus, we use the reconstruction method to address this
problem and calculate the value of $f(\mathcal{Q})$. In this method,
the Hubble parameter is known. Firstly, the gravitational field can
be described by the e-folding number
$(\mathcal{N}=\ln(\frac{a}{a_{0}}))$ and simplification of the
resulting second order differential equation allows us to deduce the
value of $f(\mathcal{Q})$. Thus, Eq.\eqref{8} in terms of e-folding
number becomes
\begin{equation}\label{9}
\mathcal{Q}(\mathcal{N})=6{H}^{2}(\mathcal{N}).
\end{equation}
To reduce the complexity, we assume the particular form of Hubble
parameter as \cite{40}
\begin{equation}\label{10}
H^{2}(\mathcal{N})=P(\mathcal{N})\Rightarrow
\mathcal{Q}(\mathcal{N})=6P(\mathcal{N}).
\end{equation}
We assume the cosmic scale factor as \cite{41}
\begin{equation}\label{11}
a(t)=(1+\mu t^{2})^{\frac{\nu}{2}},
\end{equation}
where $\mu$ and $\nu$ are positive parameters. The Hubble parameter
and  e-folding number associated to this scale factor turn out to be
\begin{equation}\label{12}
H=\frac{\dot{a}}{a}=\frac{\mu\nu t}{1+{\mu} t^2}, \quad
\mathcal{N}=-\frac{\nu}{2}\ln{(\frac{\mathcal{Q}}{6\mu{\nu}^{2}})}.
\end{equation}
Using e-folding number relation, we obtain
\begin{equation}\label{13}
H^{2}(\mathcal{N})=P(\mathcal{N})=\mu{\nu}^{2}{e}^{\frac{-2\mathcal{N}}{\mu}}.
\end{equation}
Using Eq.\eqref{10} in \eqref{7}, it follows that
\begin{equation}\label{14}
\frac{P'(\mathcal{N})}{\sqrt{P(\mathcal{N})}}-\frac{P'(\mathcal{N})
f_{\mathcal{Q}}}{\sqrt{P(\mathcal{N})}}
-\frac{12P(\mathcal{N})f_{\mathcal{Q}\mathcal{
Q}}}{\sqrt{P(\mathcal{N})}}
-\frac{f}{2}+6P(\mathcal{N})f_{\mathcal{Q}}+3P(\mathcal{N})=0.
\end{equation}
The solution of this differential equation is
\begin{eqnarray}\nonumber
f(\mathcal{Q})&=&c_1e^{\frac{\mathcal{Q}\big(-P'(\mathcal{N})-\sqrt{P'^{2}
(\mathcal{N})-24P^{3/2}(\mathcal{N})}\big)}{24P(\mathcal{N})}}+
c_2e^{\frac{\mathcal{Q}\big(-P'(\mathcal{N})+\sqrt{P'^{2}
(\mathcal{N})-24P^{3/2}(\mathcal{N})}\big)}{24P(\mathcal{N})}}
\\\label{15}
&+&2\frac{P'(\mathcal{N})+3P^{3/2}(\mathcal{N})}{\sqrt{P(\mathcal{N})}},
\end{eqnarray}
where $c_{1}$ and $c_{2}$ are the integration constants and prime is
the derivative corresponding to e-folding number. In the upcoming
sections, cosmological parameters will be discussed with the help of
this obtained $f(\mathcal{Q})$ function.

The general integral action of the $f(\mathcal{Q})$ theory is
specified as follows \cite{42}
\begin{equation}\label{16}
\mathcal{S}=\int\left(\frac{1}{2\kappa}f(\mathcal{Q})+L_{m}\right){\sqrt{-g}}
d^4x.
\end{equation}
Here, $L_{m}$ denotes the Lagrangian density associated with matter.
The associated field equations are
\begin{equation}\label{17}
\frac{2}{\sqrt{-g}}\nabla_{\xi}(f_{\mathcal{Q}}\sqrt{-g}
\mathcal{P}^{\xi}_{~\alpha\beta})+\frac{1}{2}f
g_{\alpha\beta}+f_{\mathcal{Q}}
(\mathcal{P}_{\alpha\xi\tau}\mathcal{Q}_{\beta}^{~\xi\tau}-2\mathcal{Q}^{\xi\tau}_{~~~\alpha}
\mathcal{P}_{\xi\tau\beta})=T_{\alpha\beta}.
\end{equation}
Here, $\nabla_{\xi}$ demonstrates the covariant derivative and
$T_{\alpha\beta}$ denotes the EMT. We consider the perfect fluid
configuration as
\begin{equation}\label{18}
T_{\alpha\beta}=(p_m+\rho_m)u_{\alpha}u_{\beta}+p_mg_{\alpha\beta},
\end{equation}
where $p_m$ is the pressure, $\rho_m$ is the energy density and
$u_{\alpha}$ is the four-velocity of the fluid. The resulting field
equations in the background of the flat FRW spacetime metric turn
out to be
\begin{eqnarray}\label{19}
\rho_m&=&\frac{1}{2}f-6H^{2}f',
\\\label{20}
p_m&=&2f'\dot{H}+2f''H +6f'H^{2}-\frac{1}{2}f.
\end{eqnarray}
These field equations are useful for analyzing the complex nature of
the cosmos. Banerjee et al \cite{43} examined different forms of
non-metricity to examine the mysterious universe.

\section{Bouncing Cosmology in $f(\mathcal{Q})$ Theory}

Bouncing cosmology offers a different perspective from the big bang
theory by proposing that the universe experiences periods of
contraction and expansion with a bounce. In the field of cosmology,
bouncing solutions are important as they address the initial
singularity associated with the cosmic expansion. Consequently, they
provide an alternative explanation to the big bang singularity. In a
cyclic universe, the cosmos transits from a previous contraction
phase to an expansion phase without encountering a singularity. The
cosmic bounce can be seen as a rhythmic or periodic occurrence where
the collapse of one phase leads to another cosmological event. Cai
and his colleagues \cite{44} examined the bouncing universe to
develop a non-singular bounce scenario following a contraction
phase. The following conditions must be satisfied for the viable
non-singular bouncing universe.
\begin{itemize}
\item
The scale factor must be at its minimum close to the bounce point
for a nonsingular bouncing model. The decreasing behavior of scale
factor indicates that the universe is in contracting phase, while
its increasing behavior signifies a period of cosmic expansion.
\item
The universe experiences contraction and expansion phases when
Hubble parameter is positive and negative, respectively. The bounce
point is obtained at $H=0$.
\item
The universe passes through an accelerating phase when deceleration
parameter is negative and the cosmos undergoes in an expansion era
when the deceleration parameter is positive.
\item
The EoS parameter describes the universe entering a phantom stage
when $\omega<-1$ and a quintessence phase when $-1/3<\omega<-1$.
\item
The energy density must be positive, finite as well as maximum and
pressure should be negative for the existence of non-singular
bounce.
\item
In the case of a non-singular bouncing universe, the ECs must be
violated.
\end{itemize}
These constraints have implications to our comprehension of the
early universe and serve as observable indicators of bouncing
scenarios. Therefore, these characteristics provide further insight
into the fundamental properties of the cosmos.

\subsection{Evolution of the Scale Factor}

The behavior of scale factor is crucial in understanding the
dynamics of the expanding, contracting and bouncing universe. It is
a positive function that changes over time and used to quantify the
expansion of the universe and its size. However, the cosmic time is
measured in gigayears (Gyr). The expression for the scale factor is
given in Eq.\eqref{11}. The graphical representation of scale factor
is shown in Figure \textbf{1}, which shows its decreasing nature
leading to the smallest value (non-zero) and then increasing,
providing an explanation for the bouncing pattern observed in the
initial phases of cosmic development. From this plot, it is clear
that the cosmos is shifted from a contracting phase to an expanding
phase. Furthermore, it indicates that the scale factor grows
symmetrically around the bouncing point.
\begin{figure}\center
\epsfig{file=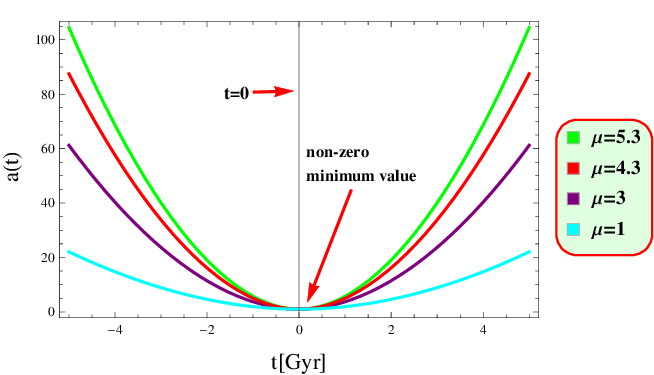,width=.5\linewidth}\caption{Behavior of scale
factor for different values of $\mu$ and $\nu=1.9$.}
\end{figure}

\subsection{Dynamics of Hubble Parameter}
\begin{figure}
\epsfig{file=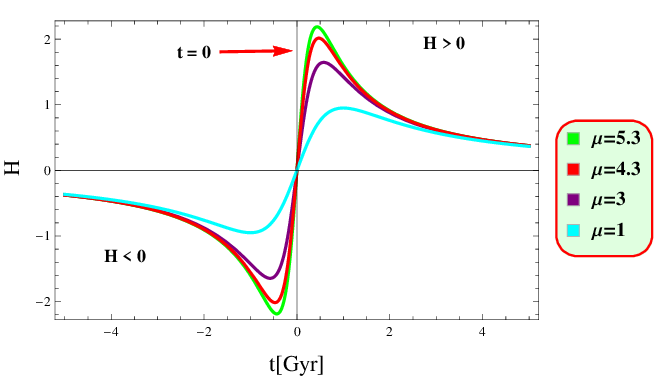,width=.5\linewidth}
\epsfig{file=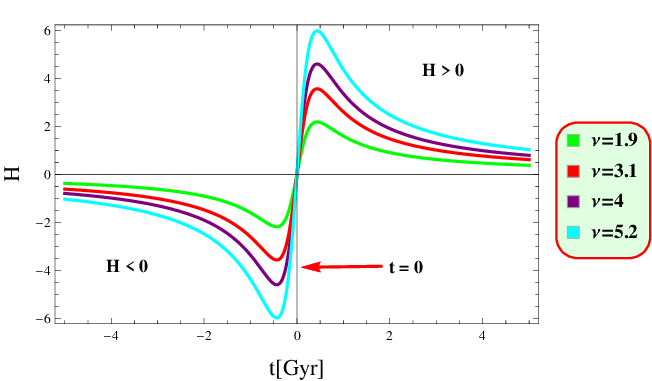,width=.5\linewidth}\caption{Behavior of Hubble
parameter for different parametric values.}
\end{figure}

Now, we use the Hubble parameter to investigate the behavior of
cosmos and other cosmic applications. The Hubble parameter
corresponding to the scale factor is shown in Eq.\eqref{12}. This
parametrization is designed to model a bouncing universe scenario,
which helps in understanding cosmic acceleration. Figure \textbf{2}
shows a transition from a contracting to an expanding universe as
$H$ becomes negative before the bounce point and positive after the
bounce for different values of $\mu$ and $\nu$. Moreover, the
negative values for model parameters do not support the cosmic
acceleration.
\begin{figure}\center
\epsfig{file=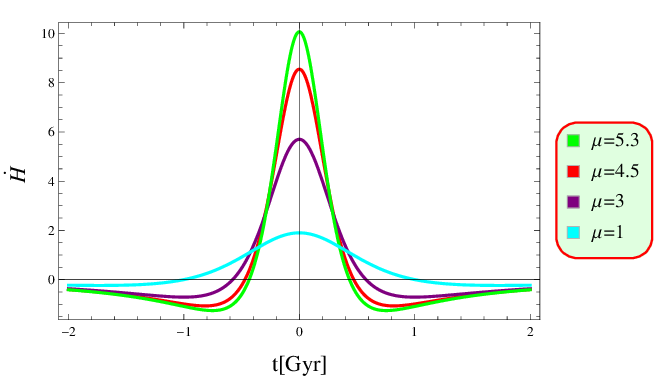,width=.5\linewidth}\caption{Evolution of
temporal derivative of Hubble parameter.}
\end{figure}

The nature of Hubble parameter is shown in Tables \textbf{1} and
\textbf{2} for various values of $\mu$ and $\nu$, respectively. The
behavior of the Hubble constant with different values of $\mu$ is
shown in Table \textbf{1} while keeping $\nu=1.9$. Table \textbf{2}
analyzes how the Hubble parameter evolves with different values of
$\nu$ by keeping $\mu$ constant at $5.3$. Figure \textbf{3} shows
that the time derivative of the Hubble parameter is positive near
the bounce point for different values of $\mu$, which demonstrates
the cosmic acceleration.
\begin{table}\caption{\textbf{Nature of Hubble Parameter for different values of
$\mu~ (\nu=1.9)$.}}
\begin{center}
\begin{tabular}{|c|c|c|c|}
\hline $\mu$ & Time Interval & Behavior of $H$ & Nature of Cosmos
\\
\hline 5.3   & $-0.5<t<0$   & $H<0$ & Contraction
\\
\hline 4.5   & $-0.5<t<0$   & $H<0$ & Contraction
\\
\hline 3     & $-0.5<t<0$   & $H<0$ & Contraction
\\
\hline 1    & $-0.5<t<0$  & $H<0$ & Contraction
\\
\hline 5.3  & $0<t<0.5$ & $H>0$ & Expansion
\\
\hline 4.5  & $0<t<0.5$ & $H>0$ & Expansion
\\
\hline 3    & $0<t<0.5$ & $H>0$ & Expansion
\\
\hline 1    & $0<t<0.5$ & $H>0$ & Expansion
\\
\hline
\end{tabular}
\end{center}
\end{table}
\begin{table}\caption{\textbf{Nature of Hubble Parameter for different values of
$\nu~(\mu=5.3)$.}}
\begin{center}
\begin{tabular}{|c|c|c|c|}
\hline $\nu$ & Time Interval & Behavior of $H$ & Nature of Cosmos
\\
\hline 1.9      & $-0.5<t<0$   & $H<0$ & Contraction
\\
\hline 3.1      & $-0.5<t<0$   & $H<0$ & Contraction
\\
\hline 4      & $-0.5<t<0$   & $H<0$ & Contraction
\\
\hline 5.2   & $-0.5<t<0$  & $H<0$ & Contraction
\\
\hline 1.9    & $0<t<0.5$ & $H>0$ & Expansion
\\
\hline 3.1   & $0<t<0.5$ & $H>0$ & Expansion
\\
\hline 4    & $0<t<0.5$ & $H>0$ & Expansion
\\
\hline 5.2    & $0<t<0.5$ & $H>0$ & Expansion
\\
\hline
\end{tabular}
\end{center}
\end{table}

\subsection{Analysis of Rate of Cosmic Expansion}

The rate of expansion is determined by the deceleration parameter,
i.e., the positive value indicates an decelerated cosmos whereas a
negative value demonstrates an accelerated universe. This parameter
is expressed as
\begin{equation}\label{21}
q=-\frac{\dot{H}}{H^{2}}-1=\frac{t^2\mu-1-t^2\mu\nu}{t^2\mu\nu}.
\end{equation}
The graphical behavior of deceleration parameter is shown in Figure
\textbf{4}, which determines that the universe is in the accelerated
expansion phase as the value of deceleration parameter is negative.
\begin{figure}\center
\epsfig{file=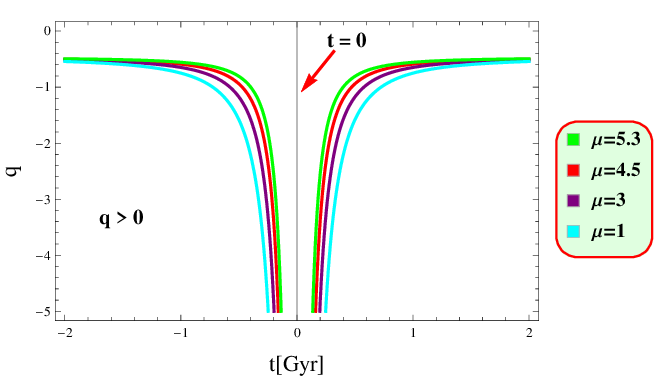,width=.5\linewidth}\caption{Nature of
deceleration parameter versus cosmic time.}
\end{figure}

\subsection{Evolution of Matter Variables}

In this subsection, we examine the impact of the reconstructed
functional form of $f(\mathcal{Q})$ given in Eq.\eqref{17} on the
dynamics of the cosmic bounce using matter variables. The
incorporation of additional terms may yield beneficial outcomes. The
reconstructed functional form provides a streamlined approach to
observe the influences of energy density and pressure on diverse
cosmic phenomena. The units for energy density and pressure are
considered as $GeV/cm^{3}$.  Using the value of model in
Eqs.$\eqref{19}$ and $\eqref{20}$, we get
\begin{eqnarray}\nonumber
\rho_m&=&\frac{1}{24(1+\mu{t}^2)^{2}\mu{\nu}^2}\bigg
[\Upsilon_1\big((1+\mu t^{2})^{2}
+12(6\mu\nu-1){\mu}^{2}{\nu}^{2}t^{2}\big)c_{1}
\\\label{22}
&+&\Upsilon_2\big((1+\mu t^{2})^{2}
-12(1+6\mu\nu){\mu}^{2}{\nu}^{2}t^{2}\big)c_{2}-72{\mu}^{2}{\nu}^{4}{t}^2\bigg],
\\\nonumber
p_m&=&\frac{1}{216(1+\mu{t}^{2})^{2}}\bigg[\frac{2\sqrt{6{\mu}}(1+\mu{t}^{2})\mu{\nu}^{2}
t}{\big(\frac{\mu\nu t^{2}}{(1+\mu
t^{2})^{2}}\big)^{\frac{3}{2}}}+648\mu{\nu}^{2}t^{2}-\frac{9}{\mu{\nu}^{2}}
\\\nonumber
&\times&\bigg[\Upsilon_1\big((1+\mu t^{2})^{2}-4(1+\mu
t^{2})(1-6\mu\nu)^{2}{\mu}^2{\nu}^{3}t+12
\\\nonumber
&\times&(-1+6\mu\nu){\mu}^{2}{\nu}^{2}
t^{2}\big)c_{1}+\Upsilon_2\big((1+\mu t^{2})^{2}-4(1+\mu t^{2})
\\\label{23}
&\times&(\mu\nu+6\mu{\nu}^{2})^{2}\mu\nu
t-12(1+6\mu\nu){\mu}^{2}{\nu}^{2} t^{2}\big)c_2\bigg]\bigg].
\end{eqnarray}
Here,
\begin{eqnarray}\nonumber
\Upsilon_1=\exp(\frac{6(1-6\mu\nu){\mu}^2{\nu}^2t^{2}}{(1+\mu
t^{2})^{2}}), \quad
\Upsilon_2=\exp(\frac{6(1+6\mu\nu){\mu}^2{\nu}^2t^{2}}{(1+\mu
t^{2})^{2}}).
\end{eqnarray}
The graphical representation of  energy density and pressure is
depicted in Figure \textbf{5}, which exhibits a positive trajectory
for energy density and negative trend for pressure. These
characteristics are crucial in understanding the dynamics of DE
models.
\begin{figure}
\epsfig{file=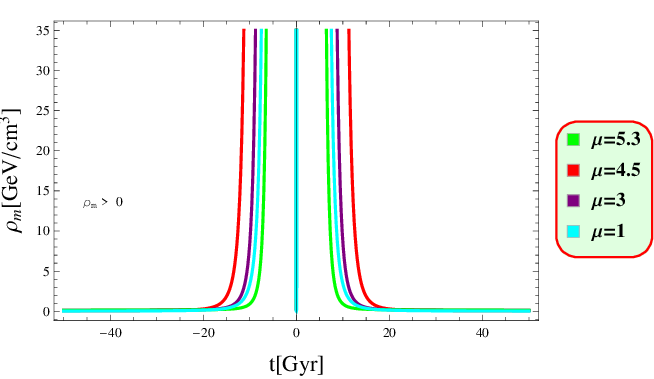,width=.5\linewidth}
\epsfig{file=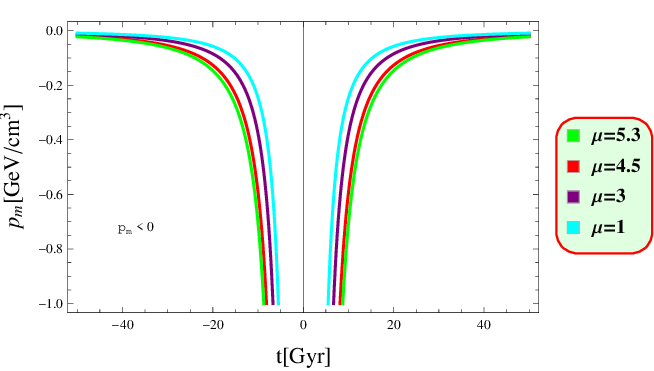,width=.5\linewidth}\caption{Profile of energy
density and pressure versus cosmic time.}
\end{figure}

\subsection{Analysis of State Parameter}

In this section, we discuss the physical attributes of different
cosmic parameters. The EoS parameter $(\omega=\frac{p}{\rho})$ can
be categorized into different stages of the cosmic development. The
matter-dominated regions such as dust, radiative fluid and stiff
matter regions are determined by $\omega=0,~\frac{1}{3},~1$,
respectively. Whereas the vacuum, phantom and quintessence cosmic
phases are represented by $\omega= -1$, $\omega <-1$ and
$-\frac{1}{3}<\omega<-1$, respectively \cite{45}. The EoS parameter
corresponding to the reconstructed $f(\mathcal{Q})$ gravity model is
obtained as
\begin{eqnarray}\nonumber
\omega&=&\frac{2}{9(\mu\nu)^{\frac{5}{2}}(1+\mu t^2)^{3}t^{3}}
\bigg[\frac{1}{\Upsilon_1}\big((1+\mu
t^{2})^{2}+12(-1+6\mu\nu){\mu}^{2}{\nu}^{2}t^{2}\big)c_{1}
\\\label{26}
&+&\frac{1}{\Upsilon_2}\big((1+\mu
t^{2})^{2}-12(1+6\mu\nu){\mu}^{2}{\nu}^{2}t^{2}\big)c_{2}
-72\mu{\nu}^{3}t^{2}\bigg].
\end{eqnarray}
The EoS parameter maintains a symmetrical behavior before and after
the bounce point, preventing the occurrence of singularities during
the bounce phases as shown in Figure \textbf{6}. This ensures a
smooth transition through the bounce phase as the EoS parameter does
not approach to infinity at any point. Furthermore, the trajectory
of the EoS parameter falls in the phantom region. This characterizes
the dynamic transformation in the nature of dark matter and DE
during the crucial phases in our universe.
\begin{figure}\center
\epsfig{file=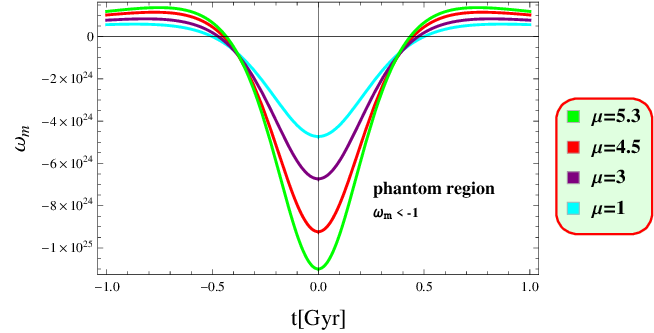,width=.5\linewidth}\caption{Plot of EoS
parameter versus cosmic time.}
\end{figure}

\subsection{Dynamics of Energy Conditions}
\begin{table}\caption{\textbf{Types of Energy Conditions}}
\begin{center}
\begin{tabular}{|c|c|}
\hline Energy Conditions & Perfect Fluid
\\
\hline NEC   & $\rho_{m}+p_{m}\geq0$
\\
\hline SECs   & $(n-3)\rho_{m}+(n-1)p_{m}$, $\rho_{m}+p_{m}\geq0$
\\
\hline DECs  &  $\rho_{m}\geq|p_{m}|$
\\
\hline  WECs  & $\rho_{m}\geq0$, $\rho_{m}+p_{m}\geq0$
\\
\hline
\end{tabular}
\end{center}
\end{table}

The ECs are essential for the evolution of geodesic structures that
are space-like, time-like, or light-like and figure out the
singularity of spacetime. These constraints are helpful to
comprehend the cosmic geometry and their relationships to the EMT.
These are classified into four categories as NEC, WECs, SECs and
DECs. Different forms of ECs are given in Table \textbf{3}. Here, we
explore the graphical representation of these energy constraints for
the reconstructed $f(\mathcal{Q})$ model. By examining these ECs, we
can determine the characteristics of cosmic geometries and their
relationship to EMT. Violation of the null energy condition yields
the violation of all ECs \cite{46}. Using Eqs.\eqref{22} and
\eqref{23}, we have
\begin{eqnarray}\nonumber
\rho_m+p_m&=&\frac{1}{24(1+\mu{t}^2)^{2}\mu{\nu}^2}\bigg
[\Upsilon_1\big((1+\mu
t^{2})^{2}+12(6\mu\nu-1){\mu}^{2}{\nu}^{2}t^{2}\big)c_{1}
\\\nonumber
&+&\Upsilon_2\big((1+\mu
t^{2})^{2}-12(1+6\mu\nu){\mu}^{2}{\nu}^{2}t^{2}\big)c_{2}-72{\mu}^{2}{\nu}^{4}{t}^2\bigg]
\\\nonumber
&+&\frac{1}{216(1+\mu{t}^{2})^{2}}
\bigg[\frac{2\sqrt{6{\mu{\nu}^{2}}}(1+\mu{t}^{2})\mu\nu
t}{\big(\frac{\mu\nu t^{2}}{(1+\mu t^{2})^{2}}\big)^{\frac{3}{2}}}
-\frac{9}{\mu{\nu}^{2}}\bigg[\Upsilon_1\big((1+\mu t^{2})^{2}
\\\nonumber
&-4&(1+\mu t^{2})(1-6\mu\nu)^{2}{\mu}^2{\nu}^{3}t
+12(-1+6\mu\nu){\mu}^{2}{\nu}^{2} t^{2}\big)c_{1}+\Upsilon_2
\\\nonumber
&\times&\big((1+\mu t^{2})^{2}-4(1+\mu t^{2})
(\mu\nu+6\mu{\nu}^{2})^{2}\mu\nu t+648\mu{\nu}^{2}t^{2}
\\\label{27}
&-&12(1+6\mu\nu){\mu}^{2}{\nu}^{2} t^{2}\big)c_2\bigg]\bigg],
\\\nonumber
\rho_m+3p_m&=&\frac{1}{24(1+\mu{t}^2)^{2}\mu{\nu}^2}\bigg
[+\Upsilon_1\big((1+\mu t^{2})^{2}+12(6\mu\nu-1)
\\\nonumber
&\times&{\mu}^{2}{\nu}^{2}t^{2}\big)c_{1}+\Upsilon_2\big((1+\mu
t^{2})^{2}-12(1+6\mu\nu){\mu}^{2}{\nu}^{2}t^{2}\big)c_{2}
\\\nonumber
&-&72{\mu}^{2}{\nu}^{4}{t}^2\bigg]+3\bigg[\frac{1}{216
(1+\mu{t}^{2})^{2}}\bigg[\frac{2\sqrt{6{\mu{\nu}^{2}}}(1+\mu{t}^{2})\mu\nu
t}{\big(\frac{\mu\nu t^{2}}{(1+\mu t^{2})^{2}}\big)^{\frac{3}{2}}}
\\\nonumber
&-&\frac{9}{\mu{\nu}^{2}}\bigg[\Upsilon_1\big((1+\mu
t^{2})^{2}-4(1+\mu t^{2})(1-6\mu\nu)^{2}{\mu}^2{\nu}^{3}t
\\\nonumber
&+&12(-1+6\mu\nu){\mu}^{2}{\nu}^{2}
t^{2}\big)c_{1}+\Upsilon_2\big((1+\mu t^{2})^{2}-4
\\\nonumber
&\times&(1+\mu t^{2})(\mu\nu+6\mu{\nu}^{2})^{2}\mu\nu
t-12(1+6\mu\nu){\mu}^{2}{\nu}^{2} t^{2}\big)c_2\bigg]
\\\label{28}
&+&648\mu{\nu}^{2}t^{2}\bigg]\bigg],
\\\nonumber
\rho_m-p_m&=&\frac{1}{24(1+\mu{t}^2)^{2}\mu{\nu}^2}\big
[-72{\mu}^{2}{\nu}^{4}{t}^2+\Upsilon_1\big((1+\mu t^{2})^{2}+12
\\\nonumber
&\times&(6\mu\nu-1){\mu}^{2}{\nu}^{2}t^{2}\big)c_{1}
+\Upsilon_2\big((1+\mu t^{2})^{2}-12(1+6\mu\nu)
\\\nonumber
&\times&{\mu}^{2}{\nu}^{2}t^{2}\big)c_{2}\bigg]
-\frac{1}{216(1+\mu{t}^{2})^{2}}\bigg[\frac{2\sqrt{6{\mu{\nu}^{2}}}(1+\mu{t}^{2})\mu\nu
t}{\big(\frac{\mu\nu t^{2}}{(1+\mu t^{2})^{2}}\big)^{\frac{3}{2}}}
\\\nonumber
&-&\frac{9}{\mu{\nu}^{2}}\bigg[\Upsilon_1\big((1+\mu
t^{2})^{2}-4(1+\mu t^{2})(1-6\mu\nu)^{2}{\mu}^2{\nu}^{3}t
\\\nonumber
&+&12(-1+6\mu\nu){\mu}^{2}{\nu}^{2}
t^{2}\big)c_{1}+\Upsilon_2\big((1+\mu t^{2})^{2}
\\\nonumber
&-&4(1+\mu t^{2})(\mu\nu+6\mu{\nu}^{2})^{2}\mu\nu
t-12(1+6\mu\nu){\mu}^{2}{\nu}^{2} t^{2}\big)\bigg]
\\\label{29}
&+&648\mu{\nu}^{2}t^{2}\bigg].
\end{eqnarray}
Figure \textbf{7} shows that there is no singularity in the vicinity
of the bouncing point as energy constraints are negative near the
bounce, providing strong indication of a violation of these ECs. In
our analysis, the violation of the ECs interpreted to achieve a
non-singular bounce that allows the universe to transition from a
contracting phase to an expanding phase without encountering a
singularity. Furthermore, it is important to note that such
violations are localized in time and are confined to the immediate
vicinity of the bounce. They do not extend to regions away from the
bounce epoch, where the DECs are restored which ensure that the
model remains physically viable in the broader context.
\begin{figure}
\epsfig{file=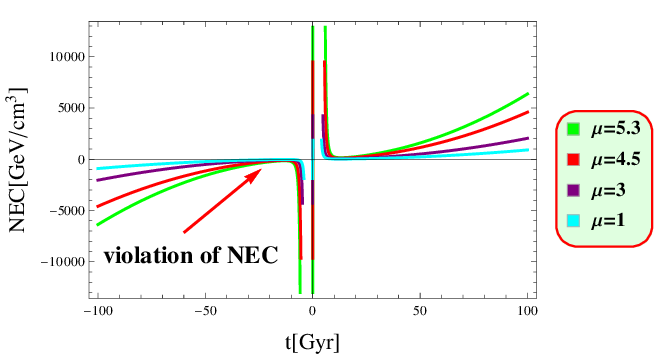,width=.5\linewidth}
\epsfig{file=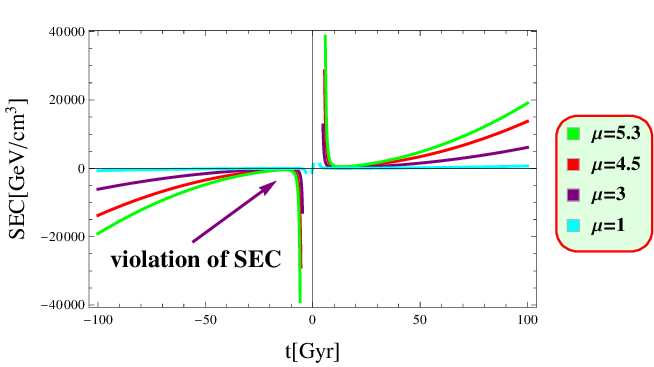,width=.5\linewidth}
\epsfig{file=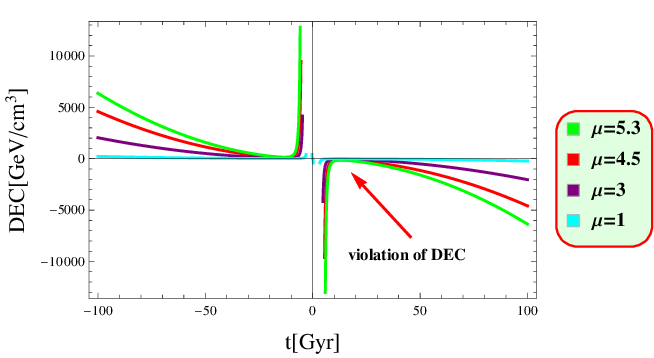,width=.5\linewidth}
\epsfig{file=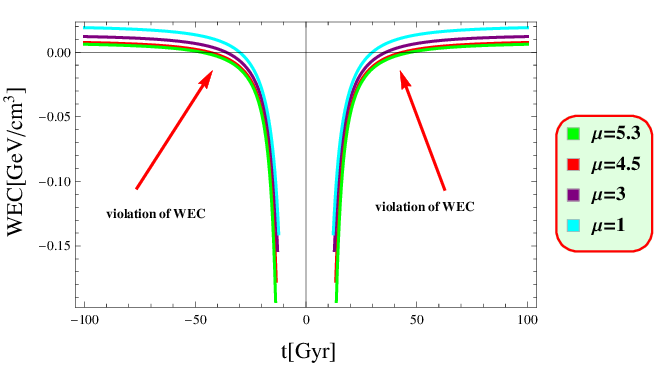,width=.5\linewidth}\caption{Behavior of energy
conditions corresponding to cosmic time.}
\end{figure}

\section{Second Law of Thermodynamics}

This analysis examines the second law of thermodynamics in the
context of $f(\mathcal{Q})$ gravity. According to this law, the
total entropy of the universe must increase over time. The total
entropy comprises the entropy of matter in the universe $(S_{in})$
and the entropy on the boundary of the horizon $(S_{on})$. The units
for entropy are Jouls per Kelvin $(JK^{-1})$. It is assumed that the
boundary of the universe is delimited by the radius of the apparent
horizon, which can be determined using the Hubble parameter
mentioned in Eq.\eqref{12}.
\begin{equation}\label{31}
R_{h}=\frac{1}{H}=\frac{\mu t^{2}+1}{t\mu\nu}.
\end{equation}
In cosmology, the entropy of the horizon is associated with the
horizon surface area as
\begin{equation}\label{32}
A= 4\pi R_{h}^{2}=\frac{4\pi(\mu
t^{2}+1)^{2}}{t^{2}{\mu}^{2}{\nu}^{2}}.
\end{equation}
By applying the Bekenstein-Hawking formulation, the entropy value at
the horizon boundary is calculated as
\begin{equation}\label{33}
S_{on}=\frac{K_{b}c^{3}}{G\hbar}\bigg[\frac{\pi(\mu
t^{2}+1)^{2}}{t^{2}{\mu}^{2}{\nu}^{2}}\bigg],
\end{equation}
where $K_{b}$ is the Boltzmann constant and
$\frac{G\hbar}{c^{3}}=L_{p}$ is the Planck's length. The time
derivative of $S_{on}$ turns out to be
\begin{equation}\label{34}
\dot{S}_{on}=\frac{2K_{b}\pi}{L_{p}{\mu}^{2}{\nu}^{2}}
\bigg({\mu}^{2}t-\frac{1}{t^{3}}\bigg).
\end{equation}
It is noted that $\dot{S}_{on}>0$ only when $\mu>\frac{1}{t^2}$.

Using the Gibbs relation, the value of entropy in the boundary of
the horizon is obtained as
\begin{equation}\label{35}
T_{h}dS_{in}=d({\rho}_{m}V)+p_{m}dV,
\end{equation}
where ${T_h}$ is the Hawking temperature on the boundary of the
horizon. The unit for Hawking temperature is considered as Kelvin
(K). The volume inside the horizon is given by
\begin{equation}\label{36}
V=\frac{4\pi}{3}\bigg(\frac{\mu t^{2}+1}{t\mu\nu}\bigg)^{3}.
\end{equation}
Differentiating Eq.\eqref{35} with respect to time $t$, we get
\begin{eqnarray}\nonumber
T_{h}\dot{S}_{in}&=&\frac{4\pi}{3}\frac{(1+\mu t^2)^3}{(\mu\nu
t)^{3}}\times\frac{t\mu}{6\mu{\nu}^{2}(1+\mu
t^{2})^2}\bigg[-36\mu{\nu}^{4}+\frac{1}{(1+\mu t^{2})^3}
\\\nonumber
&\times&\Upsilon_1\big[(1+\mu
t^2)^{4}-36{\mu}^{3}t^2(1-6\mu\nu){\nu}^{4}+216{\mu}^4{\nu}^5{t}^{2}
\\\nonumber
&\times& (1-6\mu\nu)+3\mu\big(-1-6\mu\nu+3\mu
t^2(1+2\mu\nu)\big)(\nu+\mu t^2 \nu)^2
\\\nonumber
&+&36{\mu}^{2}(\nu+\mu t^2
\nu)^3\big]c_{1}+\Upsilon_2\big[1+72{\mu}^{2}{\nu}^{5}t^2
\\\label{37} &+& \mu\big(t^2 +24{\mu}^2{\nu}^4
t+2(-3-18\mu\nu+{\mu}^2 \nu t)\big)\big]c_{2}\bigg]
\end{eqnarray}
Hawking temperature is positive and can be calculated as
\begin{eqnarray}\label{38}
T_{h}=\frac{1}{2\pi R_{h}}\big(1-\frac{\dot{R}_{h}}{2HR_{h}}\big)
=\frac{\mu\nu t}{2\pi(1+t^2\mu)}\bigg[1+\frac{1-\mu t^2}{2 \mu \nu
t^2}\bigg].
\end{eqnarray}
Since $S_{in}>0$, the total entropy should not decrease in terms of
time evolution. Thus, the second law of thermodynamics holds if the
following condition is satisfied
\begin{equation}\label{39}
\dot{S}_{tot}=\dot{S}_{on}+\dot{S}_{in}\geq0.
\end{equation}
Figure \textbf{8} shows that the total entropy of the universe
increases near the bounce over time. Furthermore, the model
indicates that the Hawking temperature is very high after the
bouncing time.
\begin{figure}
\epsfig{file=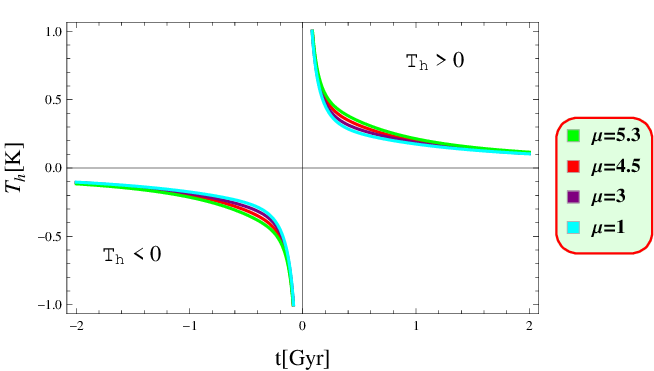,width=.5\linewidth}
\epsfig{file=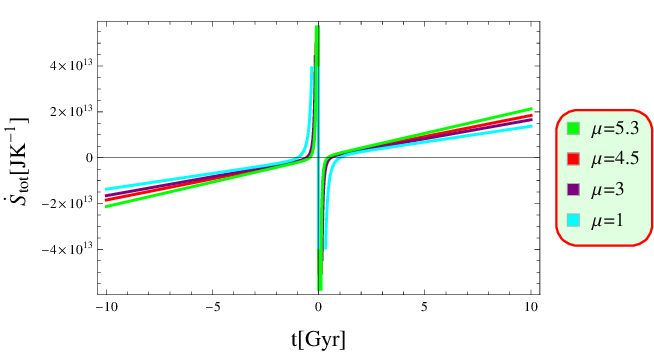,width=.5\linewidth}\caption{The evolution of
total entropy and temperature versus $t$.}
\end{figure}

\section{Conclusions}

In recent decades, researchers have faced challenges regarding the
origin and development of the universe due to the limited
observational data. As a result, cosmologists have been studying
bouncing cosmology as an alternative approach to address the
inflationary problem and singularities in the big bang model
\cite{46}. This strategy becomes relevant when dealing with
uncertainties related to early singularities. Nonsingular bouncing
solutions play a crucial role in the early universe cosmology and it
is essential to explore various facts of these solutions. The
primary aim of this research is to determine the nature of
nonsingular bounce. In this context, we have reconstructed the
functional form of $f(\mathcal{Q})$ gravity, where $\mathcal{Q}$
characterizes the gravitational interaction. The key results of our
study are outlined as follows.
\begin{itemize}
\item
The scale factor shows that the universe transits from contracting
to expansion phase as it is monotonically decreasing before the
bounce and increasing after the bounce. Also, it maintains its
minimum value before and after the bouncing spot (Figure
\textbf{1}).
\item
The Hubble parameter becomes negative prior to the bounce point,
signifying a contracting universe. This becomes zero as the cosmos
approaches to the bounce point. As the universe transitions into the
post-bounce phase, it takes on a positive value, indicating that the
cosmos undergoes the expansionary phase. (Figure \textbf{2}).
\item
Different values of the parameters $\mu$ and $\nu$ facilitate the
shift from a contraction to an expansion in the present-day universe
(Tables \textbf{1} and \textbf{2}).
\item
The time derivative of the Hubble parameter shows that the cosmos
transits from a contracting to an expanding state in the vicinity of
the bounce point, which occurs in the cosmic time interval of $-0.5$
to $0.5$  (Figure \textbf{3}).
\item
The deceleration parameter demonstrates that the universe is
undergoing an expansion phase as it is negative before and after the
bounce (Figure \textbf{4}).
\item
The positive energy density and negative pressure indicate that the
universe is currently in an expanding state (Figure \textbf{5}).
\item
The EoS parameter suggests that the universe is in the phantom era,
confirming the cosmic expansion (Figure \textbf{6}).
\item
The violation of all the ECs leads to the accelerated expansion
(Figure \textbf{7}).
\item
We have found that our reconstructed model attains high temperature
at the bouncing point and exhibits an increasing behavior of total
entropy over time (Figure \textbf{8}).
\end{itemize}

The second law of thermodynamic plays a crucial role in bouncing
cosmology and models of the accelerated expansion. Any viable
cosmological model must adhere to this thermodynamic principle. This
imposes constraints on the behavior of dark matter and DE throughout
the cycles of contraction and expansion, ensuring that entropy
consistently increases. We have explored the validity of the second
law of thermodynamics corresponding to FRW universe bounded by a
horizon. Our results indicate that the total entropy increases as
the universe approaches to the bounce point and total entropy tends
to infinity at the bounce point. We concluded an increasing behavior
of the total entropy over time, reflecting the progression towards
thermodynamic equilibrium.

Finally, we have compared our findings with $\Lambda$CDM model and
found that all ECs are violated when the bouncing requirements are
met. This behavior is not compatible with $\Lambda$CDM as in
$\Lambda$CDM only SEC is violated \cite{47}. Both the observed
accelerated expansion and the predictions of the $\Lambda$CDM model
align with this behavior \cite{48}. It is worthwhile to mention here
that the reconstruction model satisfies all the necessary conditions
for a successful non-singular stable bouncing model.

\section*{Appendix X: Computation of
$\mathcal{Q}=-\mathcal{Q}_{\xi\alpha\beta}\mathcal{P}^{\xi\alpha\beta}$}
\renewcommand{\theequation}{X\arabic{equation}}
\setcounter{equation}{0}

Non-metricity tensor is expressed as
\begin{eqnarray}\label{X1}
\mathcal{Q}&=&-g^{\alpha\beta}\big(L^{\xi}_{~\phi\alpha}
L^{\phi}_{~\beta\xi}- L^{\xi}_{~\phi\xi
}L^{\phi}_{~\alpha\beta}\big),
\end{eqnarray}
where
\begin{eqnarray}\label{X2}
L^{\xi}_{~\phi\alpha}&=&-\frac{1}{2}g^{\xi\eta}
\big(\mathcal{Q}_{\alpha\phi\eta}+
\mathcal{Q}_{\phi\eta\alpha}-\mathcal{Q}_{\eta\phi\alpha}\big),
\\\label{X3}
L^{\phi}_{~\beta\xi}&=&-\frac{1}{2}g^{\phi\tau}
\big(\mathcal{Q}_{\xi\beta\tau}+\mathcal{Q}_{\beta\xi
\tau}-\mathcal{Q}_{\tau\beta\xi}\big),
\\\label{X4}
L^{\xi}_{~\phi\xi}&=&-\frac{1}{2}g^{\xi\eta} \big(\mathcal{Q}_{\xi
\phi\eta}+ \mathcal{Q}_{\phi\eta\xi}-\mathcal{Q}_{\eta\xi
\phi}\big),
\\\label{X5}
L^{\phi}_{~\alpha\beta}&=&-\frac{1}{2}g^{\phi\tau}
\big(\mathcal{Q}_{\beta\alpha\tau}+
\mathcal{Q}_{\alpha\tau\beta}-\mathcal{Q}_{\tau\alpha\beta}\big).
\end{eqnarray}
Therefore, we get
\begin{eqnarray}\label{X6}
-g^{\alpha\beta}L^{\xi }_{~\phi\alpha}L^{\mu}_{~\beta\xi}
&=&-\frac{1}{4}\big(2\mathcal{Q}^{\xi \beta\tau}
\mathcal{Q}_{\tau\xi\beta}-\mathcal{Q}^{\xi \beta\tau}
\mathcal{Q}_{\xi\beta\tau}\big),
\\\label{X7}
g^{\alpha\beta}L^{\xi }_{~\phi\xi }L^{\phi}_{~\alpha\beta}
&=&\frac{1}{4}g^{\alpha\beta}g^{\mu \eta}\mathcal{Q}_\phi
\big(\mathcal{Q}_{\beta\alpha\eta}+\mathcal{Q}_{\alpha\eta\beta}
-\mathcal{Q}_{\eta\beta\alpha}\big),
\\\nonumber
\mathcal{Q}&=&-\frac{1}{4}\big(-\mathcal{Q}^{\xi\beta\tau}
\mathcal{Q}_{\xi\beta\tau}
+2\mathcal{Q}^{\xi\beta\tau}\mathcal{Q}_{\tau\xi \beta}
-2\mathcal{Q}^{\xi}\tilde{\mathcal{Q}}_{\xi }
\\\label{X8}
&-&2\mathcal{Q}^{\xi}\mathcal{Q}_{\xi}\big).
\end{eqnarray}
Using Eq.(\ref{20}), we have
\begin{eqnarray}\nonumber
\mathcal{P}^{\xi\alpha\beta}&=&\frac{1}{4}\bigg[-\mathcal{Q}^{\xi\alpha\beta}
+\mathcal{Q}^{\alpha\xi \beta}+\mathcal{Q}^{\beta\xi\alpha}
+\mathcal{Q}^{\xi }g^{\alpha\beta}-\tilde{\mathcal{Q}}^{\xi}
g^{\alpha\beta}- \frac{1}{2}(g^{\xi\alpha}\mathcal{Q}^{\beta}
\\\label{X9}
&+&g^{\xi \beta}\mathcal{Q}^{\alpha})\bigg],
\end{eqnarray}
\begin{eqnarray}\nonumber
-\mathcal{Q}_{\xi\alpha\beta}\mathcal{P}^{\xi\alpha\beta}
&=&-\frac{1}{4}\big(-\mathcal{Q}^{\xi\alpha\beta}
\mathcal{Q}_{\xi\alpha\beta}+2\mathcal{Q}_{\xi\alpha\beta}
\mathcal{Q}^{\alpha\xi\beta} +\mathcal{Q}^{\xi}\mathcal{Q}_{\xi}
-2\mathcal{Q}_{\xi}\tilde{\mathcal{Q}}^{\xi}\big)
\\\label{X10}
&=&\mathcal{Q}.
\end{eqnarray}

\section*{Appendix Y: Calculation of $\delta\mathcal{Q}$}
\renewcommand{\theequation}{Y\arabic{equation}}
\setcounter{equation}{0} We list all the non-metricity tensors that
will be used prior to presenting the detailed variation of the
non-metricity scalar, $\mathcal{Q}$.
\begin{eqnarray}\label{Y1}
\mathcal{Q}_{\xi\alpha\beta}&=&\nabla_{\xi}g_{\alpha\beta},
\\\label{Y2}
\mathcal{Q}^{\xi}_{~\alpha\beta}&=&g^{\xi\eta}
\mathcal{Q}_{\eta\alpha\beta}=
g^{\xi\eta}\nabla_{\eta}g_{\alpha\beta}=\nabla^{\xi}
g_{\alpha\beta},
\\\label{Y3}
\mathcal{Q}^{~~\alpha}_{\xi ~~\beta}&=&g^{\alpha\eta}
\mathcal{Q}_{\xi\eta \beta}=g^{\alpha\eta }\nabla_{\xi } g_{\eta
\beta}=-g_{\eta\beta}\nabla_{\xi }g^{\alpha\eta},
\\\label{Y4}
\mathcal{Q}^{~~\beta}_{\xi \alpha}&=&g^{\beta\eta} \mathcal{Q}_{\xi
\alpha\eta}=g^{\beta\eta}\nabla_{\xi }
g_{\alpha\eta}=-g_{\alpha\eta}\nabla_{\xi }g^{\beta\eta},
\\\label{Y5}
\mathcal{Q}^{\xi \alpha}_{~~\beta}&=&g^{\xi \eta}
g^{\alpha\tau}\nabla_{\eta}g_{\tau\beta} =g^{\alpha\tau}\nabla^{\xi
}g_{\tau\beta}=-g_{\tau\beta} \nabla^{\xi }g^{\alpha\tau},
\\\label{Y6}
\mathcal{Q}^{\xi~\beta}_{~\alpha}&=&g^{\xi \nu }
g^{\beta\tau}\nabla_{\eta }g_{\alpha\tau} =g^{\beta\tau}\nabla^{\xi
}g_{\alpha\tau}=-g_{\alpha\tau} \nabla^{\xi }g^{\beta\tau},
\\\label{Y7}
\mathcal{Q}^{~~\alpha\beta}_{\xi}&=&g^{\alpha\tau}g^{\beta
\eta}\nabla_{\xi }g_{\tau\nu }
=-g^{\alpha\tau}g_{\tau\eta}\nabla_{\xi }g^{\beta\eta }
=-\nabla_{\xi}g^{\alpha\beta},
\\\label{Y8}
\mathcal{Q}^{\xi \alpha\beta}&=&-\nabla^{\xi }g_{\alpha\beta}.
\end{eqnarray}
Using Eqs.(\ref{X6}) and (\ref{X7}), we obtain
\begin{eqnarray}\nonumber
\delta \mathcal{Q}&=&-\frac{1}{4}\delta\bigg(-\mathcal{Q}
^{\xi\beta\tau} \mathcal{Q}_{\xi \beta\tau} +2\mathcal{Q}^{\xi
\beta\tau}\mathcal{Q}_{\tau\xi\beta}
-2\mathcal{Q}^{\xi}\tilde{\mathcal{Q}}_{\xi}
+2\mathcal{Q}^{\xi}\mathcal{Q}_{\xi}\bigg),
\\\nonumber
&=&-\frac{1}{4}\bigg(-\delta \mathcal{Q}^{\xi \beta\tau}
\mathcal{Q}_{\xi \beta\tau} -\mathcal{Q}^{\xi\beta\tau}\delta
\mathcal{Q}_{\xi\beta\tau} +2\delta
\mathcal{Q}^{\xi\beta\tau}\mathcal{Q}_{\tau\xi\beta}
\\\nonumber
&+&2\mathcal{Q}^{\xi\beta\tau}\delta \mathcal{Q}_{\tau\xi\beta}
-2\delta \mathcal{Q}^{\xi}\tilde{Q}_{\xi}
-2\mathcal{Q}^{\xi}\delta\tilde{\mathcal{Q}}_{\xi} +\delta
\mathcal{Q}^{\xi}\mathcal{Q}_{\xi} +\mathcal{Q}^{\xi}\delta
\mathcal{Q}_{\xi}\bigg),
\\\nonumber
&=&-\frac{1}{4}\bigg[\mathcal{Q}_{\xi\beta\tau} \nabla^{\xi}\delta
g^{\beta\tau}-\mathcal{Q}^{\xi\beta\tau} \nabla_{\xi}\delta
g_{\beta\tau}-2\mathcal{Q}_{\tau\xi\beta} \nabla^{\xi}\delta
g^{\beta\tau}+2\mathcal{Q}^{\xi\beta\tau}\nabla_{\tau}\delta
g_{\xi\beta}\\\nonumber
&+&2\tilde{\mathcal{Q}}_{\tau}\nabla^{\tau}g^{\alpha\beta}\delta
g_{\alpha\beta}+2\tilde{\mathcal{Q}}_{\tau}g_{\alpha\beta}\nabla^{\tau}\delta
g^{\alpha\beta}-2\mathcal{Q}^{\tau}\nabla^{\nu }\delta
g_{\tau\eta}-\mathcal{Q}_{\tau}\nabla^{\tau}g^{\alpha\beta}\delta
g_{\alpha\beta}\\
\label{Y9} &-&\mathcal{Q}_{\tau}g_{\alpha\beta}\nabla^{\tau}\delta
g^{\alpha\beta}-\mathcal{Q}^{\tau}\nabla_{\tau}g^{\alpha\beta}\delta
g_{\alpha\beta}-\mathcal{Q}^{\tau}g_{\alpha\beta}\nabla_{\tau}\delta
g^{\alpha\beta}\bigg].
\end{eqnarray}
Here, We use the following equations as
\begin{eqnarray}\label{Y10}
\delta g_{\alpha\beta}&=&-g_{\alpha\xi}\delta g^{\xi\nu
}g_{\eta\beta},
\\\nonumber
-\mathcal{Q}^{\xi\beta\tau}\nabla_{\xi}\delta
g_{\beta\tau}&=&-\mathcal{Q}^{\xi\beta\tau}\nabla_{\xi}\big(-g_{\beta\eta}\delta
g^{\nu \vartheta}g_{\vartheta\tau}\big)
\\\label{Y11}
&=&2\mathcal{Q}^{\xi
\psi}_{~~\beta}\mathcal{Q}_{\xi\psi\alpha}\delta
g^{\alpha\beta}+\mathcal{Q}_{\xi
\beta\tau}\nabla^{\xi}g^{\beta\tau},
\\\label{Y12}
2\mathcal{Q}^{\xi \beta\tau}\nabla_{\tau}\delta
g_{\xi\beta}&=&-4\mathcal{Q}^{~~\psi\tau}_{\alpha}\mathcal{Q}_{\tau\psi\beta}\delta
g^{\alpha\beta}-2\mathcal{Q}_{\beta\tau\xi}\nabla^{\xi}g^{\beta\tau},
\\\nonumber
-2\mathcal{Q}^{\tau}\nabla^{\eta}\delta
g_{\tau\eta}&=&2\mathcal{Q}^{\xi}\mathcal{Q}_{\beta\xi\alpha}\delta
g^{\alpha\beta}+2\mathcal{Q}_{\alpha}\tilde{\mathcal{Q}}_{\beta}\delta
g^{\alpha\beta}
\\\label{Y13}
&+&2\mathcal{Q}_{\beta}g_{\xi\tau}\nabla^{\xi}g^{\beta\tau}.
\end{eqnarray}
Thus, Eq.(\ref{Y9}) becomes
\begin{equation}\label{Y14}
\delta \mathcal{Q}=2\mathcal{P}_{\xi \beta\tau}\nabla^{\xi}\delta
g^{\beta\tau}-\big(\mathcal{P}_{\alpha\xi\eta}\mathcal{Q}^{~~\xi\eta}_{\beta}
-2\mathcal{Q}^{\xi\eta}_{~~\alpha}\mathcal{P}_{\alpha\eta\beta}\big)\delta
g^{\alpha\beta},
\end{equation}
where
\begin{eqnarray}\nonumber
&&2\mathcal{P}_{\xi\beta\tau}=-\frac{1}{4}\bigg[2\mathcal{Q}_{\xi\beta\tau}-
2\mathcal{Q}_{\tau\xi\beta}-2\mathcal{Q}_{\beta\tau\xi}+2
\mathcal{Q}_{\beta}g_{\xi\tau}+2\tilde{\mathcal{Q}}_{\xi}g_{\beta\tau}
\\\label{Y15}
&&-2\mathcal{Q}_{\xi})g_{\beta\tau}\bigg],
\\\nonumber
&&4\big(\mathcal{P}_{\alpha\xi\eta}\mathcal{Q}^{~~\xi\eta}_{\beta}
-2\mathcal{Q}^{\xi\eta}_{~~\alpha}\mathcal{P}_{\alpha\eta\beta}\big)=
2\mathcal{Q}^{\xi\eta}_{~~\beta}\mathcal{Q}_{\xi\eta\alpha}-4
\mathcal{Q}^{~~\xi \eta}_{\alpha}\mathcal{Q}_{\eta\xi\beta}
\\\label{Y16}
&&+2\tilde{\mathcal{Q}}^{\xi}\mathcal{Q}_{\xi\alpha\beta}
+2\mathcal{Q}^{\xi}\mathcal{Q}_{\beta\xi\alpha}
+2\mathcal{Q}_{\alpha}\tilde{\mathcal{Q}}_{\beta}-
\mathcal{Q}^{\xi}\mathcal{Q}_{\xi\alpha\beta}.
\end{eqnarray}

\section*{Appendix Z: Calculation of $\mathcal{Q}=6H^{2}$}
\renewcommand{\theequation}{Z\arabic{equation}}
\setcounter{equation}{0} From Eq.(\ref{X10}), we have
\begin{equation}\label{Z1}
\mathcal{Q}=-\frac{1}{4}\big(-\mathcal{Q}_{\xi\alpha\beta}
\mathcal{Q}^{\xi\alpha\beta}
+2\mathcal{Q}_{\xi\alpha\beta}\mathcal{Q}^{\alpha\xi\beta}
+\mathcal{Q}_{\xi}\mathcal{Q}^{\xi}-2\mathcal{Q}_{\xi}
\tilde{\mathcal{Q}^{\xi}}\big).
\end{equation}
For the case of FRW metric, we get
\begin{eqnarray}\label{Z2}
-\mathcal{Q}_{\xi\alpha\beta}\mathcal{Q}^{\xi\alpha\beta}
&=&\nabla_{\xi}g_{\alpha\beta}
\nabla^{\xi}g^{\alpha\beta}=4(3H^{2}),
\\\label{Z3}
\mathcal{Q}_{\xi\alpha\beta}\mathcal{Q}^{\xi\alpha\beta}
&=&\nabla_{\xi}g_{\alpha\beta}
\nabla^{\xi}g^{\alpha\beta}=-4(3H^{2}),
\\\label{Z4}
\mathcal{Q}_{\xi}\mathcal{Q}^{\xi}&=&(g_{\tau\alpha}
\nabla_{\xi}g^{\tau\alpha}) (g_{\psi\beta}\nabla^{\xi
}g^{\psi\beta})=-4(3H)^{2},
\\\label{Z5}
\mathcal{Q}_{\xi}\tilde{\mathcal{Q}^{\xi}}
&=&(g_{\alpha\tau}\nabla_{\xi}g^{\alpha\tau})
(\nabla_{\eta}g^{\xi\eta})=-4(T^{2}+3H^{2}).
\end{eqnarray}
Using Eqs.\eqref{Z2} -\eqref{Z5} in \eqref{Z1}, we obtain
\begin{equation}\label{Z6}
\mathcal{Q}=-\frac{1}{4}\bigg[12H^{2}-36H^{2}\bigg]=6H^{2}.
\end{equation}
\\
\textbf{Data Availability Statement:} No data was used for the
research described in this paper.

\end{document}